\documentclass[aps,pra,reprint,superscriptaddress,nobibnotes]{revtex4-1}

\usepackage{graphicx}
\usepackage{bm}

\usepackage{color}

\renewcommand{\vec}[1]{{\mathbf #1}}

\newcommand{\ket}[1]{|#1\rangle}

\begin{document}

\title{Magnetic-field-driven localization of light in a {cold-atom gas}}

\author{S.E. Skipetrov}
\email[]{Sergey.Skipetrov@lpmmc.cnrs.fr}
\affiliation{Universit\'{e} Grenoble Alpes, LPMMC, F-38000 Grenoble, France}
\affiliation{CNRS, LPMMC, F-38000 Grenoble, France}

\author{I.M. Sokolov}
\email[]{ims@is12093.spb.edu}
\affiliation{Department of Theoretical Physics, State Polytechnic University, 195251 St. Petersburg, Russia}

\date{\today}

\begin{abstract}
We discover a transition from extended to localized quasi-modes for light in a gas of immobile two-level atoms in a magnetic field. The transition takes place either upon increasing the number density of atoms in a  {strong} field or upon increasing the field at a high enough density. It has many characteristic features of a disorder-driven (Anderson) transition but is strongly influenced by near-field interactions between atoms and the anisotropy of the atomic medium induced by the magnetic field.
\end{abstract}

\maketitle

The transition from extended to localized eigenstates upon increasing disorder in a quantum or wave system is called after Philip Anderson who was the first to predict it for electrons in disordered solids \cite{anderson58}.  {More recently}, this transition was studied for various types of quantum particles (cold atoms \cite{chabe08}, Bose-Einstein condensates \cite{jendr12}) as well as for classical waves (light \cite{wiersma97,storzer06,sperling13}, ultrasound \cite{hu08,aubry14}). In the most common case of time-reversal symmetric systems  {invariant under spin rotation} Anderson transition takes place for a three-dimensional (3D) disorder only, eigenstates of low-dimensional systems being always localized \cite{abrahams79,evers08}. Anderson localization of light may find applications in the design of future quantum-information devices \cite{sapienza10}, miniature lasers \cite{liu14} and solar cells \cite{pratesi13}. However, no undisputable experimental observation of optical Anderson transition in 3D exists to date since alternative explanations were proposed for all published reports of it \cite{scheffold99,beek12,scheffold13}. Moreover, we have recently shown that the simplest theoretical model in which light is scattered by point scatterers (atoms) does not predict Anderson localization of light at all \cite{skip14}.

In the present Letter we show that an external magnetic field may induce a transition between extended and localized states for light in a gas of cold atoms. Magnetic field is an important and unique means of controlling wave propagation in disordered media. On the one hand, it breaks down the time-reversal invariance leading to a suppression of weak localization in electronic \cite{bergmann84} and optical \cite{lenke2000} systems and to metal-insulator transitions in topological insulators \cite{delplace12}. On the other hand, by profoundly modifying the scattering properties of individual scatterers the magnetic field can produce an enhancement of the coherent backscattering peak for light scattered by atoms with a degenerate {\em ground} state \cite{sigwarth04,sigwarth13}. Our work adds a new element in the mosaic of magnetic-field-induced phenomena in disordered systems by demonstrating that the removal of degeneracy of the {\em excited} atomic state due to the Zeeman effect and the resulting reduction of the strength of resonant dipole-dipole interactions between nearby atoms \cite{afrousheh06} are sufficient to induce a transition from extended to localized states in a dense atomic system where Anderson localization does not take place in the absence of the field \cite{skip14}. This critical phenomenon stands out from other magneto-optical effects that take place in disordered media (including also the photonic Hall effect \cite{rikken96} and Hanle effect in coherent backscattering \cite{labeyrie02}) which only give rise to weak corrections to wave transport.

We consider an ensemble of $N \gg 1$ identical two-level atoms at random position $\{ \vec{r}_i \}$ inside a spherical volume $V$ of radius $R$. The resonant frequency $\omega_0$ of atoms defines the natural length scale $1/k_0 = c/\omega_0$, where $c$ is the vacuum speed of light. The ground state $\ket{g_i}$ of an isolated atom $i$ is nondegenerate with the total angular momentum $J_g = 0$, whereas the excited states $\ket{e_i}$ is three-fold degenerate with $J_e = 1$. The three degenerate substates $\ket{e_{im}}$ correspond to the three possible projections $m = 0$, $\pm 1$ of the total angular momentum $\vec{J}_e$ on the quantization axis $z$. The natural lifetime $1/\Gamma_0$ of the excited state sets the time scale of the problem. The atoms are subject to a uniform magnetic field $\vec{B} \parallel z$ and interact with the free electromagnetic field surrounding them. The system ``atoms $+$ field'' is described by the following Hamiltonian \cite{cohen92, morice95, sigwarth13}:
\begin{eqnarray}
{\hat H} &=& \sum\limits_{i=1}^{N} \sum\limits_{m=-1}^{1} \hbar \omega_0 | e_{im} \rangle
\langle e_{im}| +
\sum\limits_{\mathbf{s} \perp \mathbf{k}} \hbar ck
\left( {\hat a}_{\mathbf{k} \mathbf{s}}^{\dagger} {\hat a}_{\mathbf{k}\mathbf{s}} + \frac12 \right)
\nonumber \\
&-& \sum\limits_{i=1}^{N} {\hat \mathbf{D}}_i \cdot {\hat \mathbf{E}}(\mathbf{r}_i) + \frac{1}{2 \epsilon_0}
\sum\limits_{i \ne j}^{N} {\hat \mathbf{D}}_i \cdot {\hat \mathbf{D}}_j \delta(\mathbf{r}_i - \mathbf{r}_j)
\nonumber \\
&+&  {g_e} \mu_B \vec{B} \cdot \vec{J}_e.
\label{ham}
\end{eqnarray}
Here $\hbar$ is the Planck's constant divided by $2 \pi$, $\vec{k}$ and $\vec{s}$ are the wave and the polarization vectors of the modes of the free electromagnetic field, ${\hat a}_{\mathbf{k} \mathbf{s}}^{\dagger}$ (${\hat a}_{\mathbf{k}\mathbf{s}}$) are the corresponding creation (annihilation) operators, ${\hat \mathbf{D}}_i$ are the atomic dipole operators, ${\hat \mathbf{E}}(\mathbf{r}_i)$ is the electric displacement vector divided by the vacuum permittivity $\epsilon_0$, $\mu_B$ is the Bohr magneton,  {and $g_e$ is the Land\'{e} factor of the excited state.}

Previous work \cite{fofanov11,sokolov11} demonstrated that in the absence of magnetic field ($\vec{B} = 0$), the degrees of freedom corresponding to the electromagnetic field can be traced out leading to an effective Hamiltonian describing the dynamics of $N$ atoms coupled by the electromagnetic field. This effective Hamiltonian takes the form of a $3N \times 3N$ Green's matrix $G$ describing the propagation of light between the atoms \cite{skip14}. The same approach can be used when $\vec{B} \ne 0$ leading to the following Green's matrix:
\begin{eqnarray}
G_{e_{i m} e_{j m'}} &=& \left( \mathrm{i} -  {2 m \Delta} \right) \delta_{e_{i m} e_{j m'}} -
\frac{2}{\hbar \Gamma_0} (1 - \delta_{e_{i m} e_{j m'}})
\nonumber \\
&\times&
\sum\limits_{\mu, \nu}
{d}_{e_{i m} g_i}^{\mu} {d}_{g_j e_{j m'}}^{\nu}
\frac{e^{\mathrm{i} k_0 r_{ij}}}{r_{ij}^3}
\nonumber
\\
&\times& \left\{
\vphantom{\frac{r_{ij}^{\mu} r_{ij}^{\nu}}{r_{ij}^2}}
 \delta_{\mu \nu}
\left[ 1 - \mathrm{i} k_0 r_{ij} - (k_0 r_{ij})^2 \right]
\right.
\nonumber \\
&-&\left. \frac{r_{ij}^{\mu} r_{ij}^{\nu}}{r_{ij}^2}
\left[3 - 3 \mathrm{i} k_0 r_{ij} - (k_0 r_{ij})^2 \right]
\right\},
\label{green}
\end{eqnarray}
where $\Delta = g_e \mu_B B/\hbar\Gamma_0$ is the Zeeman shift in units of the natural line width, $\vec{d}_{e_{i m} g_i} = \langle J_e m|{\hat \mathbf{D}}_i | J_g 0 \rangle$, and
$\vec{r}_{ij} = \vec{r}_i - \vec{r}_j$.

\begin{figure}
\hspace*{-5mm}
\includegraphics[width=1.1\columnwidth]{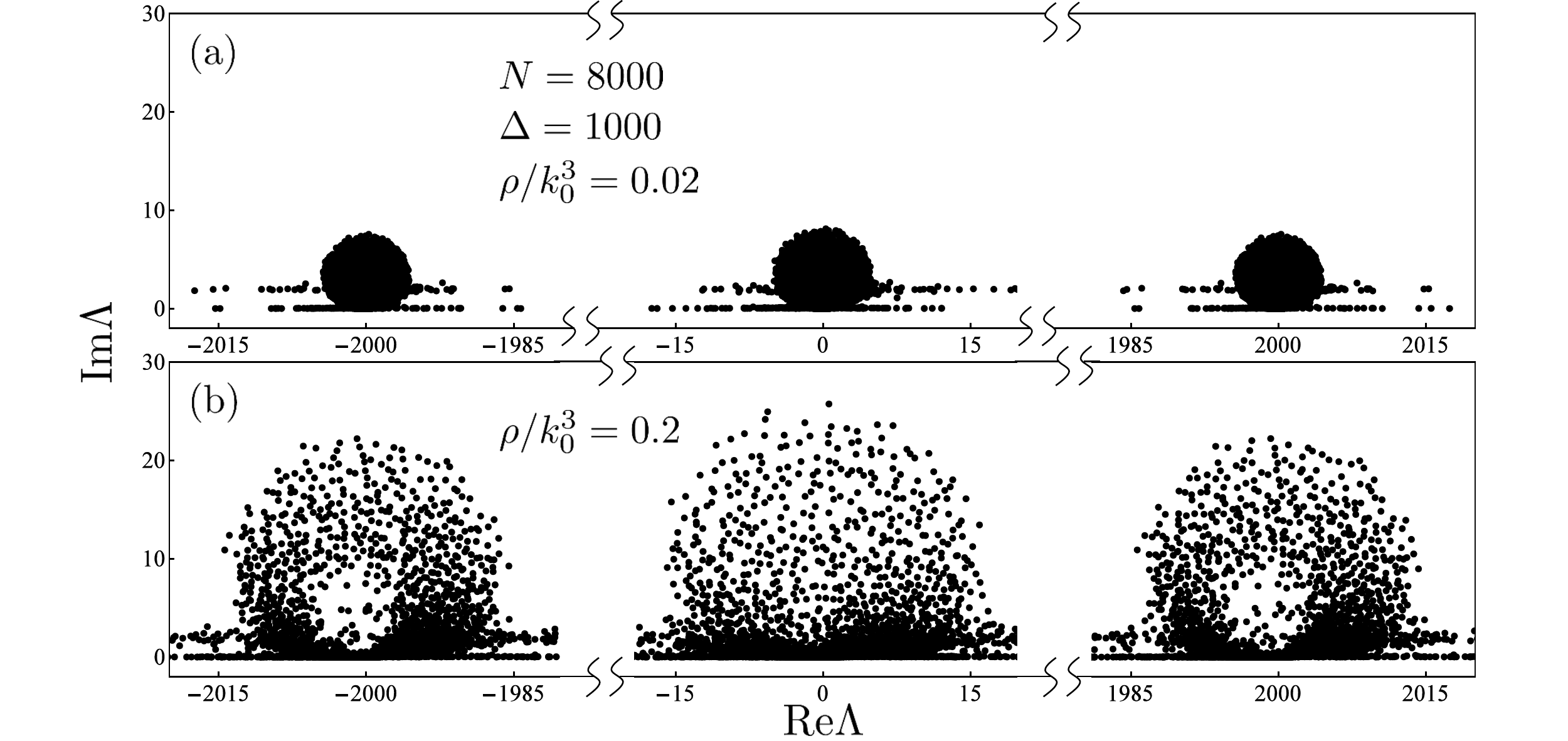}
\caption{Complex eigenvalues $\Lambda$ of a representative random realization of the Green's matrix for $N = 8 \times 10^3$ two-level atoms in a strong magnetic field ($\Delta = 10^3$) at low (a) and high (b) densities of atoms.}
\label{fig:ev}
\end{figure}

In the absence of magnetic field ($\vec{B} = 0$) the eigenvalues of the Green's matrix $G$ concentrate in a roughly circular domain on the complex plane roughly symmetric with respect to the vertical axis $\mathrm{Re} \Lambda = 0$ and almost touching the horizontal axis $\mathrm{Im} \Lambda = 0$ \cite{skip14,bellando14}. The field splits the eigenvalues into three equal groups centered around  $\mathrm{Re} \Lambda = -2 m \Delta$ ($m = 0$, $\pm 1$) \cite{pin04}, see Fig.\ \ref{fig:ev}. The three groups of eigenvalues become well separated in the limit of strong magnetic field $\Delta \gg 1$ to which we will restrict our consideration in the present Letter. Although at a low density $\rho = N/V$ the three groups of eigenvalues are similar [Fig.\ \ref{fig:ev}(a)], the groups corresponding to $m = \pm 1$ start to differ significantly from the $m = 0$ group at higher densities [Fig.\ \ref{fig:ev}(b)]. In particular, the $m = \pm 1$ groups of eigenvalues develop ``holes'' that were previously associated with Anderson localization in the framework of the scalar model of wave scattering \cite{skip11,goetschy11}.

\begin{figure*}
\includegraphics[width=0.9\textwidth]{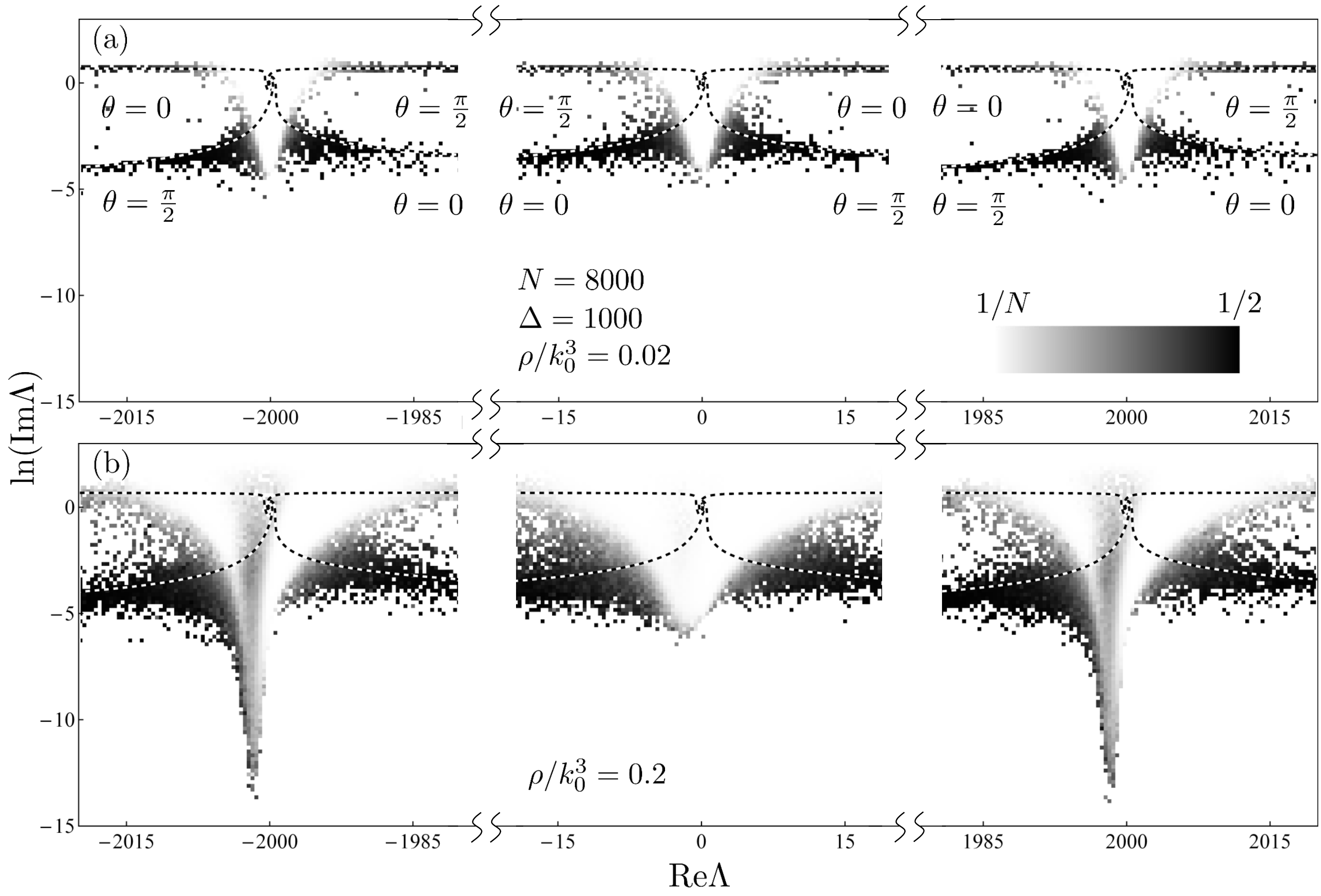}
\caption{Gray-scale maps of the average IPR at low (a) and high (b) densities of atoms in a strong magnetic field  {$\Delta = 10^3$}. Dashed lines show lines along which the eigenvalues of a two-atom system would be situated for atoms placed along the direction of magnetic field ($\theta = 0$) or perpendicular to it ($\theta = \frac{\pi}{2}$). For $0 < \theta < \frac{\pi}{2}$ the corresponding eigenvalues are in between the two lines. The gray level of each small square in the figure is obtained by averaging over IPR of all eigenvalues that fall inside the square for 16 different realizations of the random Green's matrix.}
\label{fig:ipr}
\end{figure*}

To see whether localized states indeed appear at high densities of atoms, we analyze the inverse participation ratios (IPRs) of eigenvectors $\bm{\psi}_n$ of the Green's matrix $G$,  $\mathrm{IPR}_n = \sum_{i=1}^N |\psi_{n e_i}|^4/$ $(\sum_{i=1}^N |\psi_{n e_i}|^2)^2$, where $|\psi_{n e_i}|^2 = \sum_{m = -1}^1 (\psi_{n e_i})_m^2$ is the square of the length of the vector $\psi_{n e_i} = \{ (\psi_{n e_i})_m \}$. Low IPR $\sim 1/N$ corresponds to an extended state whereas IPR $\sim 1/M > 1/N$ signals a state localized on $M < N$ atoms. Figure \ref{fig:ipr}(a) shows that at low density of atoms most of the eigenvectors have low IPRs with the eigenvectors localized on pairs of closely located atoms being an exception. These ``subradiant'' states exist at any density and should be distinguished from localized states that are due to the multiple scattering of light on many atoms and that appear at higher densities in relatively narrow bands of frequencies $\mathrm{Re} \Lambda$ on the left from the resonances $\mathrm{Re} \Lambda = \pm 2 \Delta$ [see Fig.\ \ref{fig:ipr}(b)]. These states may have smaller IPRs than the subradiant states but they have significantly longer lifetimes (i.e. smaller $\mathrm{Im} \Lambda$).

The appearance of states localized on large clusters of atoms in a magnetic field is due to the removal of degeneracy of the excited states $\ket{e_i}$ by the field. As a result, the transitions $\ket{g_i} \to \ket{e_{im}}$ effectively decouple for different $m$ since photons scattered on these transitions have frequencies discrepant by  $\approx 2 g_e \mu_B B/\hbar \gg \Gamma_0$. As a consequence, a behavior similar to the scalar case may be expected for a given $m$ with, in particular, localized states appearing at high densities of atoms as found in the scalar model \cite{skip14}. However, as follows from Fig.\ \ref{fig:ipr}, this naive picture is largely oversimplified because it does not explain the absence of localized states near $\mathrm{Re} \Lambda = 0$ corresponding to $m = 0$. A more detailed study shows that indeed, the full vector problem can be reduced to an effective scalar one in the limit of strong magnetic field, but the effective Green's matrix following from this analysis is different from the one corresponding to scalar waves. We have found that for {$\Delta \gg 1$}, the group of eigenvalues corresponding to a given $m$ can be approximately found by diagonalizing the effective $N \times N$ Green's matrix
\begin{eqnarray}
G_{ij} &=& \left( \mathrm{i} -  {2 m \Delta} \right) \delta_{ij}
+ (1- \delta_{ij}) \frac{e^{\mathrm{i} k_0 r_{ij}}}{k_0 r_{ij}}
\nonumber \\
&\times&
 {\left\{ \vphantom{\frac{\mathrm{i}}{k_0 r_{ij}}} c_m \left[ 1 - (-1)^m \cos^2 \theta \right] \right.}
\nonumber \\
&+&  {\left. c_m (-1)^m \left[ \frac{\mathrm{i}}{k_0 r_{ij}} -
\frac{1}{(k_0 r_{ij})^2} \right] (1 - 3 \cos^2 \theta) \right\}},\;\;\;\;
\label{eff}
\end{eqnarray}
where  {$c_m = (3/8)[3 + (-1)^m]$ and} $\theta$ is the angle between $\vec{r}_{ij}$ and the $z$ axis.

Equation (\ref{eff}) explains the differences between $m = 0$ and $m = \pm 1$ seen in Figs.\ \ref{fig:ev} and \ref{fig:ipr}. First, the far-field contribution to $G_{ij}$ given by the second line of Eq.\ (\ref{eff}) varies from 0 to 1 for $m = 0$ and from $\frac12$ to 1 for $m = \pm 1$ as a function of $\theta$. It is thus  {closer} to its scalar-wave value of 1 in the former case, suggesting that the case of $m = \pm 1$ may be {better approximated by} the scalar  {model} than the case of $m = 0$. Second, the near-field term [the third line of Eq.\ (\ref{eff})] is a factor of two smaller for $m = \pm 1$ than for $m = 0$. Because near-field terms responsible for resonant dipole-dipole interactions between nearby atoms suppress light scattering \cite{balik13,pellegrino14} and prevent Anderson localization \cite{skip14}, their weakness for $m = \pm 1$ is an advantage. We see therefore that both far- and near-field features of Eq.\ (\ref{eff}) are closer to its scalar approximation for $m = \pm 1$ than for $m = 0$. This explains the appearance of localized states for $m = \pm 1$ rather than for $m = 0$ transitions.

\begin{figure*}
\includegraphics[width=0.9\textwidth]{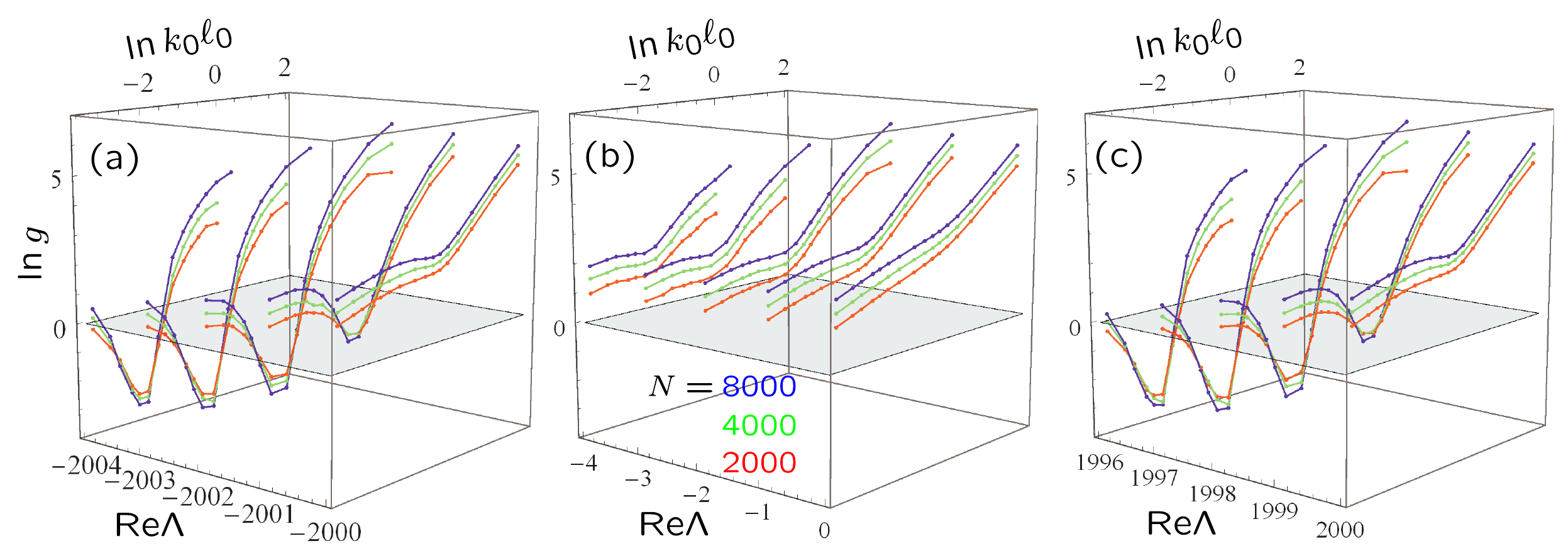}
\caption{Thouless number $g$ as a function of the bare Ioffe-Regel parameter $k_0 \ell_0 = k_0^3/6\pi\rho$ for a strong magnetic field  {$\Delta = 10^3$}. The curves are obtained by averaging over a unit interval of $\mathrm{Re} \Lambda$ around their positions and over 50, 25 or 16 realizations of random positions of $N$ atoms for $N = 2000$, 4000 and 8000, respectively. Different curves at the same value of $\mathrm{Re} \Lambda$ correspond to different numbers of atoms $N$. The gray plane corresponds to $g = 1$.}
\label{fig:3d}
\end{figure*}

\begin{figure*}
\includegraphics[width=0.9\textwidth]{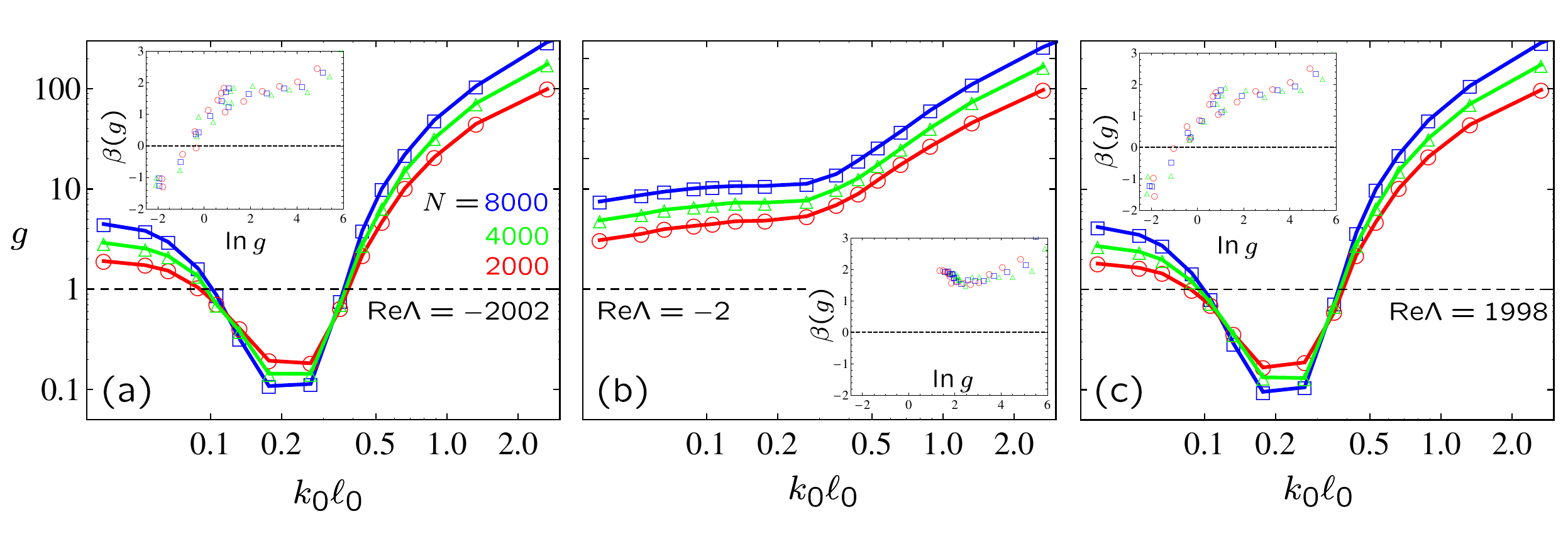}
\caption{Same as Fig.\ \ref{fig:3d} but for selected values of $\mathrm{Re} \Lambda = -2002$(a), $-2$(b) and 1998(c). The insets show the scaling function $\beta(g)$ estimated from the numerical data of the main plots.}
\label{fig:g}
\end{figure*}

To have a quantitative characterization of the localization transition demonstrated in Fig.\ \ref{fig:ipr}, we compute the Thouless number $g = \delta \omega/\Delta \omega$ that we define as a ratio of the inverse of the average lifetime of eigenstates $\delta \omega = \langle 1/\mathrm{Im} \Lambda \rangle^{-1}$ to the average eigenvalue spacing along the horizontal axis $\Delta \omega = \langle \mathrm{Re} \Lambda_n - \mathrm{Re} \Lambda_{n-1} \rangle$ \cite{abrahams79,skip14,wang11}.
At a given $\mathrm{Re} \Lambda$, $g$  reaches small values $g < 1$  {expected for localized states} only at large densities corresponding to $k_0 \ell_0 = k_0^3/6\pi\rho < 1$ and only for $\mathrm{Re} \Lambda$ corresponding to $m = \pm 1$ (see Fig.\ \ref{fig:3d}), in agreement with Fig.\ \ref{fig:ipr}.
The independence of $g$ from the sample size at the points where curves corresponding to the same value of $\mathrm{Re} \Lambda$ but different $N$ cross---a hallmark of critical behavior---is further illustrated in Fig.\ \ref{fig:g} where we reproduce $g(k_0 \ell_0)$ for $\mathrm{Re} \Lambda$ slightly shifted to the left of the single-atom resonances $\mathrm{Re} \Lambda = -2 m \Delta$.  The localization transition is also evidenced by the scaling function $\beta(g) = \partial \ln g/\partial \ln k_0 R$ \cite{abrahams79} shown in the insets of Fig.\ \ref{fig:g}. $\beta(g)$ changes sign for $\mathrm{Re} \Lambda = -2002$ and 1998 but not for $\mathrm{Re} \Lambda = -2$ proving that the {localization} transition takes place at large frequency shifts $\mathrm{Re} \Lambda \approx \pm 2 \Delta$ but not around the fundamental resonance $\mathrm{Re} \Lambda = 0$.

The localization transition reported here takes place under conditions when not only the disorder-induced multiple scattering of photons is strong but cooperative effects leading to Dicke super- and sub-radiance \cite{gross1982} and resonant dipole-dipole interactions between neighboring atoms \cite{balik13,pellegrino14} are important as well. Therefore, despite its overall similarity with the Anderson transition (see also Ref.\ \cite{suppl}), it remains to be seen if this transition can be classified as such. Nonetheless, for light of a given frequency $\omega = \omega_0 - (\Gamma_0/2) \mathrm{Re} \Lambda$  the localized regime $g \lesssim 1$  is realized only in the intermediate range of $k_0 \ell_0$ (e.g., $k_0 \ell_0 \approx 0.1$--$0.4$ for $\mathrm{Re} \Lambda = -2002$ in Fig.\ \ref{fig:g}) which corresponds to large sizes of the atomic cloud $k_0 R > 1$ (e.g., $k_0 R \approx 15$--$24$ at $N = 8000$). Hence, the localized states disappear in the Dicke limit $k_0 R < 1$ when the cooperative effects dominate. This suggests that the latter are not the main driving force behind the reported localization transition.

In conclusion, we have found that the magnetic field can induce a transition from extended to localized states for light in an ensemble of identical immobile two-level atoms. This is due to the removal of degeneracy of the excited atomic state by the field and the resulting partial suppression of resonant dipole-dipole interactions between nearby atoms.
Our theoretical predictions can be directly verified in experiments with, e.g., Sr atoms that have a nondegenerate ground state and were already used to study multiple scattering of light \cite{bidel02}. Theoretical analysis of light scattering in dense clouds of alkali atoms (such as, e.g., Rb$^{85}$) is, however, much more involved \cite{pellegrino14,sheremet12} and our results cannot be trivially extended to this case.

SES thanks the Agence Nationale de la Recherche for financial support under grant ANR-14-CE26-0032 LOVE.






\renewcommand{\vec}[1]{{\mathbf #1}}
\renewcommand{\thefigure}{S\arabic{figure}}
\renewcommand{\theequation}{S\arabic{equation}}

\bibliographystyle{apsrev4-1}
\renewcommand*{\citenumfont}[1]{S#1}
\renewcommand*{\bibnumfmt}[1]{[S#1]}

\setcounter{equation}{0}
\setcounter{figure}{0}


\newpage

\onecolumngrid

\begin{center}
\Large{\bf{Supplemental Material}}
\end{center}




\begin{center}
\parbox{0.75\textwidth}{\small We present additional evidence for the localization transition reported in the main text. In particular, we provide more details about the evolution of IPR maps shown in Fig.\ 2 with the number density of atoms, analyze the minimum value of the decay rate of quasi-modes, and study the suppression of eigenvalue repulsion due to the appearance of localized quasi-modes. In addition, we discuss some of the subtle details of light scattering in magnetic field and show that ``escape channels'' that appear in the atomic medium in the magnetic field due to photons with certain combinations of polarization and propagation direction do not shorten the lifetime of excited atomic states.}
\end{center}

\twocolumngrid


The primary purpose of this Supplemental Material is to characterize the localization transition reported in the main text in more detail and to show that, in many aspects, it exhibits the behavior expected for the Anderson transition driven by disorder (section \ref{evidence}). We start by analyzing the density dependence of IPR of eigenvectors of the Green's matrix (2) complementing the results presented in Fig.\ 2 (section \ref{iprmaps}). Then we study the behavior of the minimum decay rate of quasi-modes (section \ref{decay}). Finally, the statistics of nearest eigenvalue spacings and the phenomenon of eigenvalue repulsion are addressed in section \ref{repulsion}. In section \ref{remarks} we discuss some of the subtleties of light scattering by atoms in a magnetic field and, in particular, show by analyzing emission and absorption diagrams of individual atoms that photons that can propagate in the atomic medium without scattering due to a special relation between their polarization and propagation direction do not influence the lifetime of the excited atomic states.

\section{Evidence for the localization transition}
\label{evidence}

\subsection{Inverse participation ratio}
\label{iprmaps}

\begin{figure*}
\includegraphics[width=0.84\textwidth]{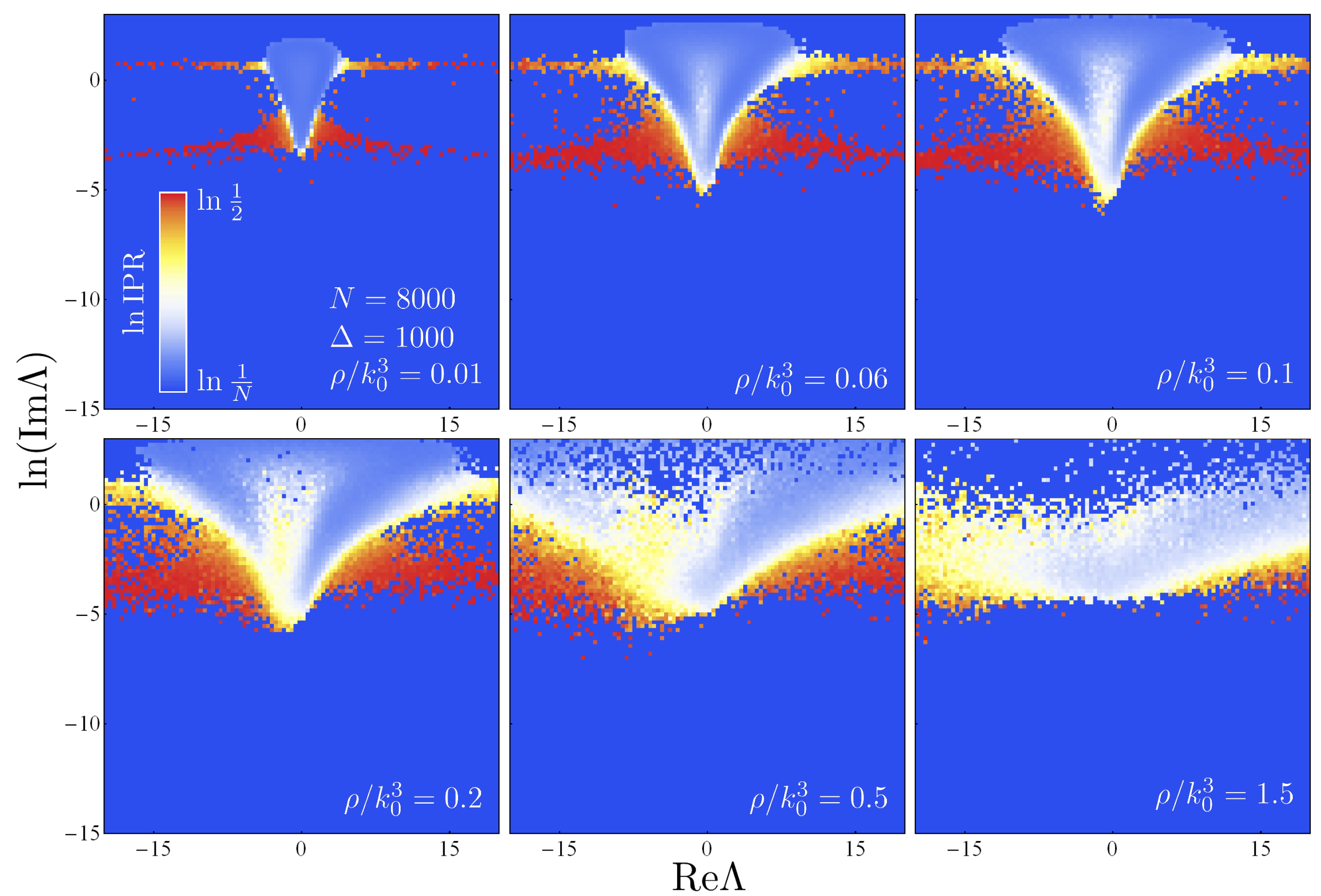}
\caption{False color maps of the logarithm of average IPR at 6 different densities for the part of spectrum near $\mathrm{Re} \Lambda = 0$. The color of each small square in the figure is obtained by taking the logarithm of the IPR averaged over all eigenvalues that fall inside the square for 6 different realizations of the random Green's matrix. The deep blue color corresponds to the parts of the complex plane where $\mathrm{IPR} \leq 1/N$ or where no eigenvalues were found.}
\label{figiprcenter}
\end{figure*}

\begin{figure*}
\includegraphics[width=0.84\textwidth]{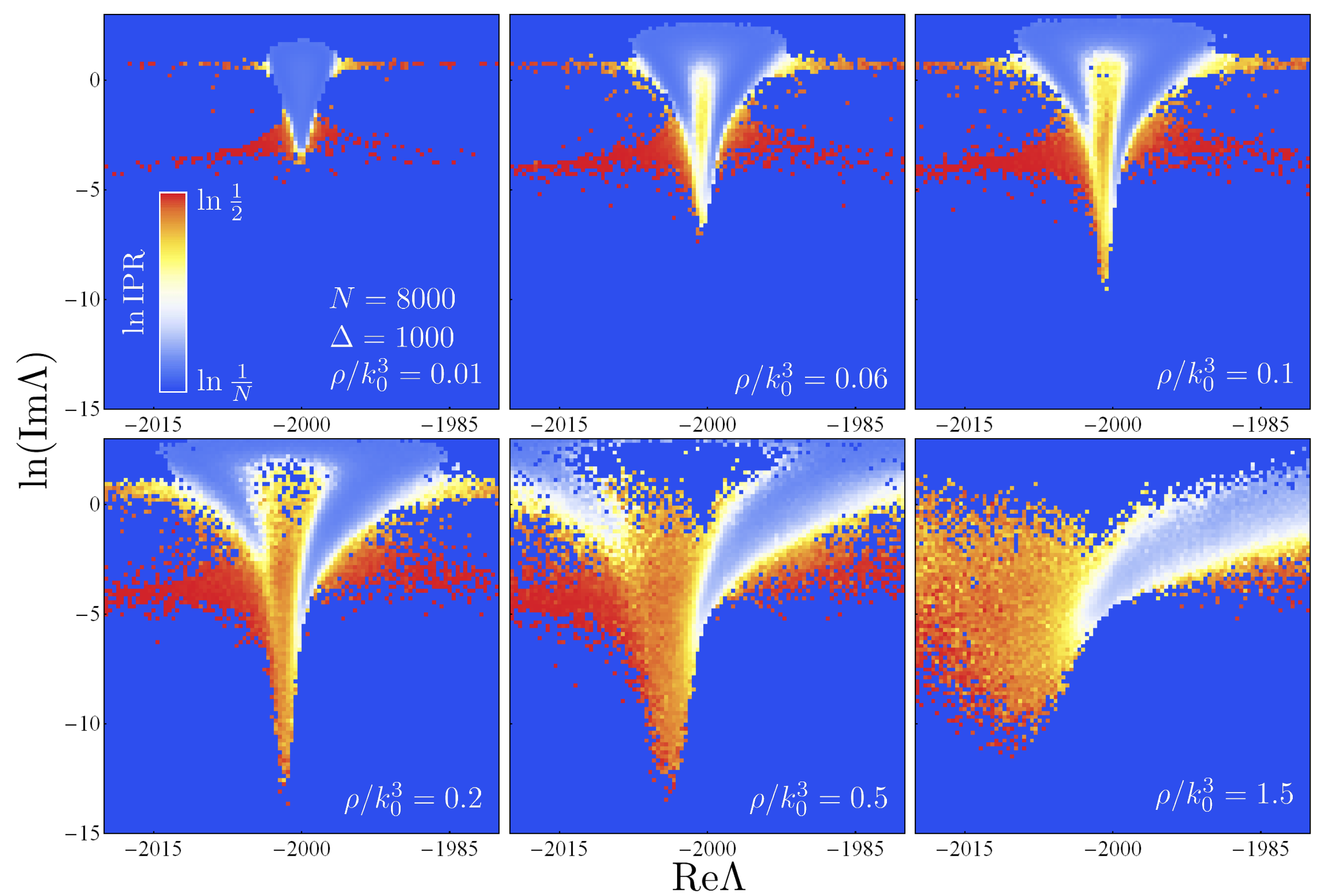}
\caption{Same as Fig.\ \ref{figiprcenter} but for the part of spectrum near $\mathrm{Re} \Lambda = -2000$.}
\label{figiprleft}
\end{figure*}

It is instructive to analyze how IPR (see the main text) of eigenvectors of the Green's matrix (2) evolves when the number density of atoms increases \cite{foot1}.
In addition to gray-scale IPR maps presented in Fig.\ 2, we show false color maps of the logarithm of average IPR at several additional densities in Figs.\ \ref{figiprcenter} and \ref{figiprleft}. The color scale of these figures allows distinguishing between the subradiant states localized on pairs of closely located atoms that appear in red and states localized on larger clusters of atoms that show up in other colors from the orange to the light blue.
Figure\ \ref{figiprcenter} clearly shows that for the states that correspond to eigenvalues with $\mathrm{Re} \Lambda$ close to 0 (the cloud of eigenvalues corresponding to $m = 0$, see the main text), IPR maps do not change qualitatively when the density increases from $\rho/k_0^3 = 0.01$ to 1.5. The cloud of eigenvalues grows in size, states localized on pairs of atoms ($\mathrm{IPR} = \frac12$) show up along the two-atom subradiant branches shown by dashed lines in Fig.\ 2, but no localized states with very small decay rates $\mathrm{Im} \Lambda$ appear upon increasing the density. States with significant IPRs in the middle of the eigenvalue cloud, near $\mathrm{Re} \Lambda \simeq -2$ (visible in white and yellow for $\rho/k_0^3 > 0.1$) are not sufficient to reach $g < 1$ and, most importantly, do not show critical behavior [the curves $g(k_0 \ell_0)$ corresponding to different $N$ do not cross and $\beta(g)$ does not change sign in Figs.\ 3 and 4]. In contrast, for the eigenvalues with $\mathrm{Re} \Lambda$ close to $\pm 2 \Delta$, IPR maps change qualitatively with increasing density and a band of localized states with very small decay rates $\mathrm{Im} \Lambda$ appears for densities $\rho/k_0^3 \gtrsim 0.1$ (see Fig.\ \ref{figiprleft} for $m = 1$, a similar scenario takes place around $\mathrm{Re} \Lambda = 2000$ for $m = -1$). The band widens and starts to disappear by merging with the two-atom subradiant branch for $\rho/k_0^3 \gtrsim 1$.

Figures\ \ref{figiprcenter} and \ref{figiprleft} clearly demonstrate the advantage of analyzing properties of eigenvectors of the Green's matrix as a function of both the real and imaginary parts of the corresponding eigenvalues instead of projecting on one of the axes ($\mathrm{Re} \Lambda$ or $\mathrm{Im} \Lambda$). The two-atom subradiant states typically have large frequency shifts with respect to single-atom resonances $\mathrm{Re} \Lambda = -2 m \Delta$ whereas the nontrivial states localized on larger clusters of atoms appear for near-resonant $\mathrm{Re} \Lambda$. Hence, the two types of states are clearly separated on the complex plane and can be distinguished. This advantage is lost when a projection on one of the axes, and, in particular, on the imaginary axis, is performed. In this case one observes a dependence that is due to both the two-atom subradiant states and the states localized on larger atomic clusters. The two cannot be disentangled anymore.

Before closing up the discussion of IPR maps shown in Figs.\ \ref{figiprcenter} and \ref{figiprleft} we would like to attract the attention of the reader to the fact that the correspondence between small decay rates $\mathrm{Im} \Lambda$ and the localized nature of the corresponding states should be used with care. Indeed, even if it might seem natural that localized states should have small decay rates and, vice versa, that small decay rates are likely to correspond to states localized in space, we see from Figs.\ \ref{figiprcenter} and \ref{figiprleft} that states with a significant IPR (e.g., white squares) systematically appear at large $\mathrm{Im} \Lambda$ and even at $\mathrm{Im} \Lambda > 1$  which corresponds to lifetimes which are even shorter than the lifetime of the excited state of an isolated atom. Therefore, making any conclusions about the spatial structure of eigenvectors of the Green's matrix based uniquely on the analysis of eigenvalues $\Lambda$ (as it was done in Refs.\ \cite{pin04x, akker08x, bellando14x}, for example) is dangerous and should necessarily be supported by the analysis of eigenvectors themselves, similarly to the analysis that we present in Fig.\ 2 and Figs.\ \ref{figiprcenter} and \ref{figiprleft}.

\subsection{Minimum decay rate}
\label{decay}

\begin{figure}
\includegraphics[width=0.9\columnwidth]{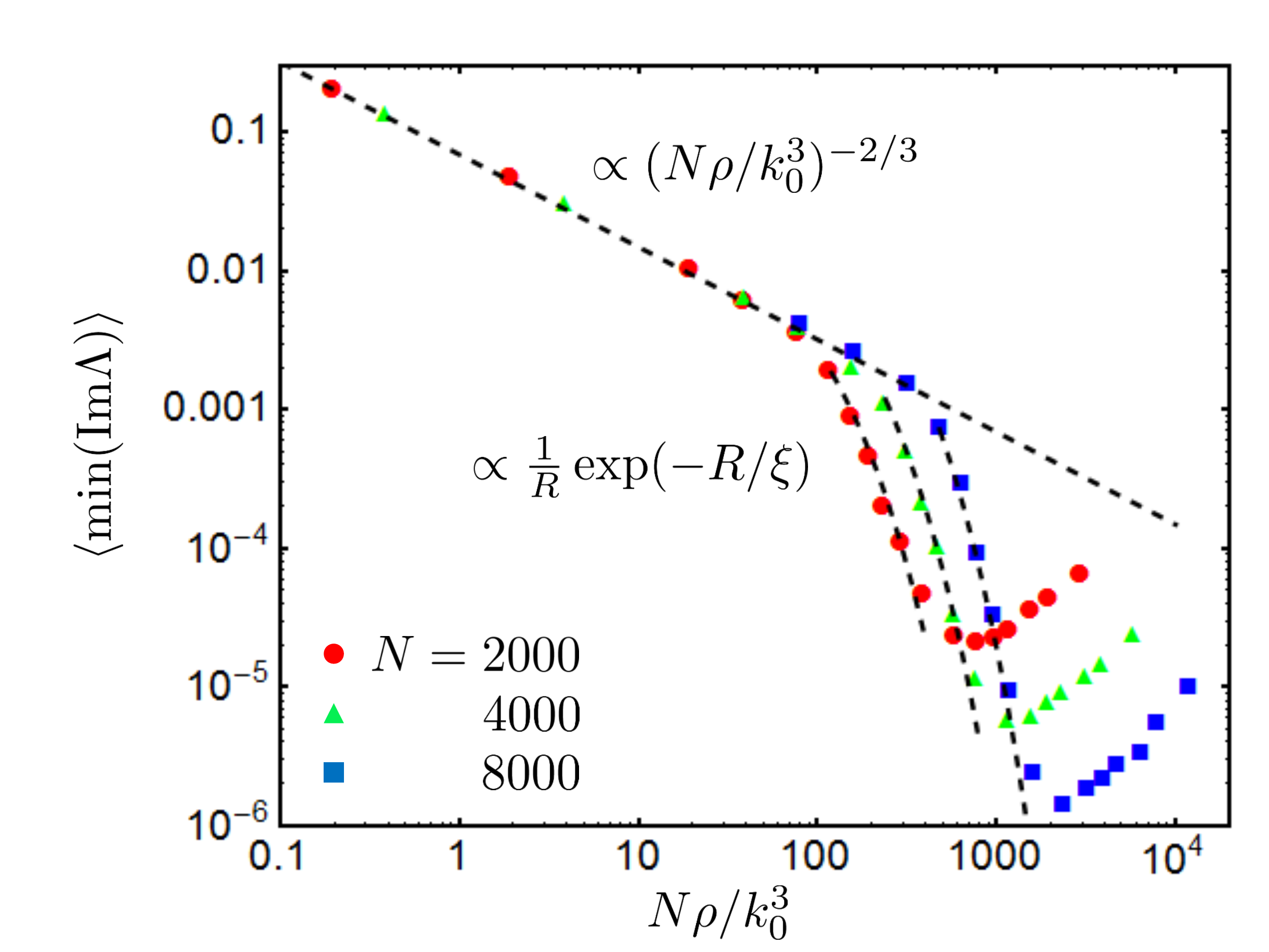}
\caption{Average value of the minimum value of the dimensionless decay rate $\mathrm{Im} \Lambda$ shown as a function of the number of atoms $N$ multiplied by the dimensionless atomic density $\rho/k_0^3$ in a strong magnetic field ($\Delta = 10^3$) for three different $N$. The numerical data (symbols) are averaged over 50, 25 and 18 realizations for $N = 2000$, 4000 and 8000, respectively.  The dashed straight line shows the expected result (\ref{min1}) in the low-density regime. The faster decay of $\langle \min(\mathrm{Im} \Lambda) \rangle$ at higher densities is fitted to Eq.\ (\ref{min2}) as explained in the text.}
\label{figminim}
\end{figure}

Let us first remind that the complex eigenvalues $\Lambda_n$ of the Green's matrix defined by Eq.\ (2) of the main text yield eigenfrequencies $\omega_n = \omega_0 -(\Gamma_0/2) \mathrm{Re} \Lambda_n$ and decay rates $\Gamma_n/2 = (\Gamma_0/2) \mathrm{Im} \Lambda_n$ of the quasi-modes $\psi_n$ of the atomic system. The temporal evolution of the quasi-mode $\psi_n$ obeys $\psi_n \propto \exp[-i \omega_n t - (\Gamma_n/2) t]$. The probability distribution $p(\mathrm{Im} \Lambda)$ of dimensionless decay rates $\mathrm{Im} \Lambda$ contains important information about the nature of quasi-modes in the system although it cannot be considered as the only proof of any statement concerning the spatial structure of the modes (extended, localized, etc., see the discussion at the end of the previous subsection). However, once the existence of spatially localized modes is demonstrated by studying the eigenvectors of the Green's matrix (see Fig. 2 and Figs.\ \ref{figiprcenter} and \ref{figiprleft}), for a mode $\psi_n$ that is weakly coupled to the exterior of the disordered sample one can assume $\Gamma_n \propto |\psi_n(\vec{r}_s)|^2$, where $\vec{r}_s$ is the ``typical'' point at the open surface of the medium \cite{kottos05x}. Because for a mode localized deep inside a disordered sample $r_s \sim R$, where $R$ is the size of the sample, the exponential decrease of $\Gamma_n$ with $R$ may be considered as a signature of exponential localization of the mode $\psi_n$ in space.

In order to focus our attention on the modes that have the smallest decay rates, we analyze the average value of the minimum dimensionless decay rate $\mathrm{Im} \Lambda$. In the absence of magnetic field, a scaling
\begin{eqnarray}
\langle \min(\mathrm{Im} \Lambda) \rangle \propto (N \rho/k_0^3)^{-2/3}
\label{min1}
\end{eqnarray}
was derived for this quantity in the low-density regime in the scalar approximation \cite{skip11ax} and in the full vector model \cite{goetschy11cx}. This scaling was further confirmed by independent numerical simulations \cite{bellando14x}. Moreover, Eq.\ (\ref{min1}) that is due to subradiant states localized on pairs of closely located atoms, was shown to hold at any density for the vector case \cite{bellando14x}. In the presence of a strong magnetic field, however, Fig.\ \ref{figminim} shows a breakdown of Eq.\ (\ref{min1}) at sufficiently high densities $\rho/k_0^3 \gtrsim 0.1$. In the region $0.06 \lesssim \rho k_0^3 \lesssim 0.2$ the numerical data for all $N$ can be fit by a phenomenological formula
\begin{eqnarray}
\langle \min(\mathrm{Im} \Lambda) \rangle \propto \frac{1}{R} \exp\left[ -\frac{R}{\xi(\rho)} \right]
\label{min2}
\end{eqnarray}
with a density-dependent localization length $\xi(\rho) \propto |\rho - \rho_c|^{-\nu}$. Figure \ref{figminim} shows fits obtained with reasonable values $\rho_c/k_0^3 = 0.041$ and $\nu = 1.11$. It is important to note that these fits do not allow determining the critical density $\rho_c$ and the critical exponent $\nu$ with any acceptable precision because fits of similar quality can be obtained for a range of values $\rho_c/k_0^3 = 10^{-2}$--$10^{-1}$ and $\nu = 1$--$2$. They witness, however, that our numerical data are compatible with the exponential decay of $\langle \min(\mathrm{Im} \Lambda) \rangle$ with sample size $R$ beyond a certain critical value of density, which is one of the signatures of exponential localization of eigenstates expected for the disorder-induced (Anderson) localization mechanism.

\begin{figure}
\includegraphics[width=0.9\columnwidth]{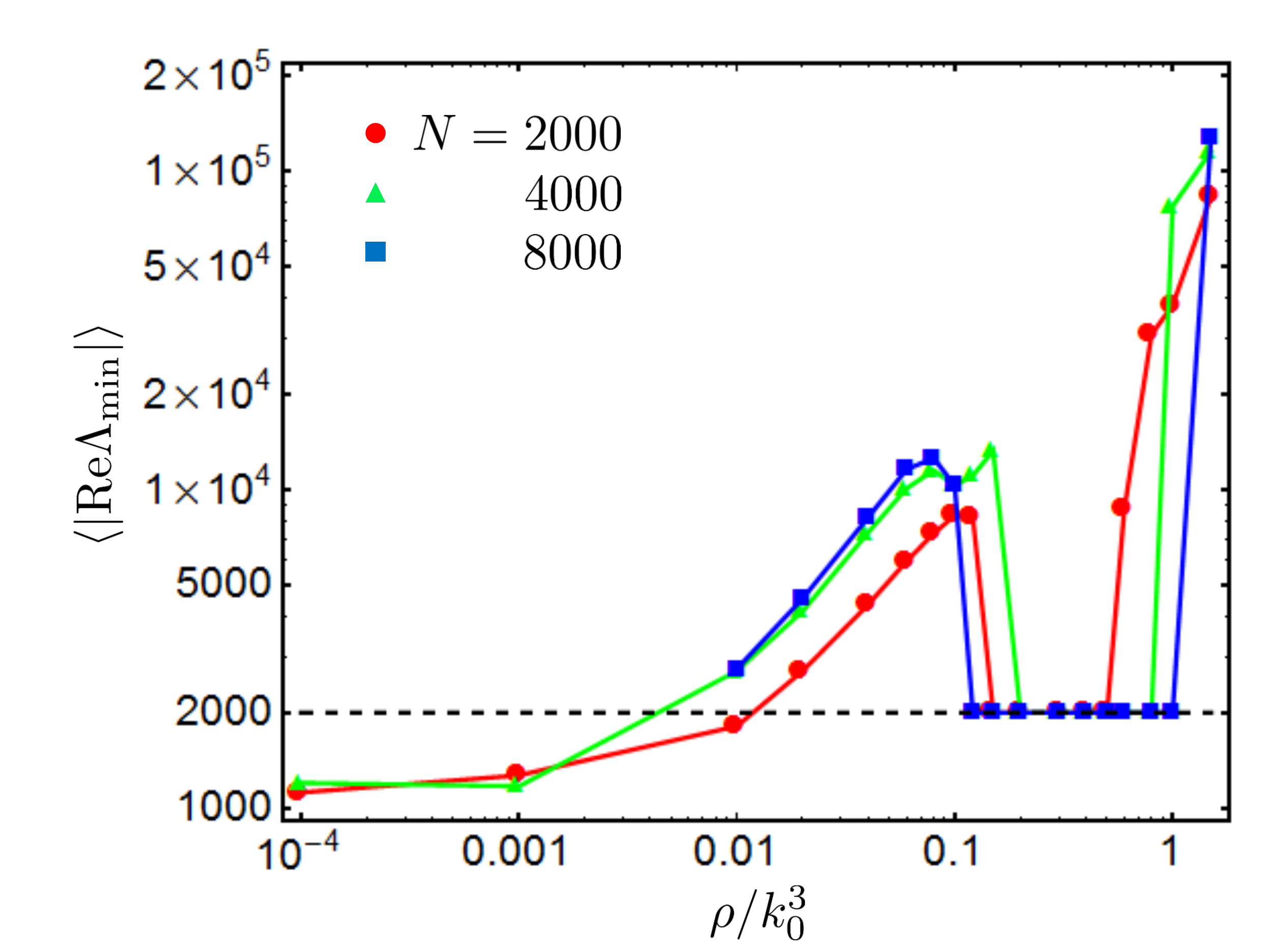}
\caption{Average absolute value of the real part of $\Lambda$ at which the minimum of $\mathrm{Im} \Lambda$ shown in Fig.\ \ref{figminim} is attained, as a function of the dimensionless atomic density $\rho/k_0^3$ in a strong magnetic field ($\Delta = 10^3$) for three different $N$. The numerical data (symbols) are averaged over 50, 25 and 18 realizations for $N = 2000$, 4000 and 8000, respectively.  The dashed horizontal line shows $\langle |\mathrm{Re} \Lambda_{\mathrm{min}}| \rangle = 2 \Delta$ expected in the intermediate density regime in which nontrivial localized states appear in the system.}
\label{figminimre}
\end{figure}

It is also instructive to see at which value of $\mathrm{Re} \Lambda$ the minimum values of $\mathrm{Im} \Lambda$ shown in Fig.\ \ref{figminim} are attained. As we show in Fig.\ \ref{figminimre}, the average value of $|\mathrm{Re} \Lambda_{\mathrm{min}}|$ at which $\mathrm{Im} \Lambda$ is minimized first grows with density to very large values $\langle |\mathrm{Re} \Lambda_{\mathrm{min}}| \rangle \sim 10^4$ before abruptly dropping to an $N$-independent value $\langle |\mathrm{Re} \Lambda_{\mathrm{min}}| \rangle = 2 \Delta$ at $\rho/k_0^3 \sim 0.1$. The fast initial growth of $\langle |\mathrm{Re} \Lambda_{\mathrm{min}}| \rangle$ with density takes place in the regime in which $\langle \min(\mathrm{Im} \Lambda) \rangle$ is due to subradiant states localized on pairs of closely located atoms and decays as a power-law with $N \rho/k_0^3$ (see Fig.\ \ref{figminim}). These subradiant states typically have large frequency shifts corresponding to large values of $|\mathrm{Re} \Lambda|$. However, we clearly see that for $\rho/k_0^3 \gtrsim 0.1$ the nature of states that have minimum decay rates changes abruptly. Now $\langle \min(\mathrm{Im} \Lambda) \rangle$ is dominated by the states localized on larger clusters of atoms (see Fig. 2 and Figs.\ \ref{figiprcenter} and \ref{figiprleft}) that have frequency shifts $\mathrm{Re} \Lambda \approx \pm 2 \Delta$ \cite{foot2}.
This situation is preserved until a sufficiently high density ($\rho/k_0^3 \sim 1$ in Fig.\ \ref{figminimre}) at which the localized states at $\mathrm{Re} \Lambda \approx \pm 2 \Delta$ start to disappear and $\langle \min(\mathrm{Im} \Lambda) \rangle$ becomes dominated by the same subradiant states as in the low-density limit. This change of regime is also witnessed by $\langle \min(\mathrm{Im} \Lambda) \rangle$ that grows with density and approaches the line $\langle \min(\mathrm{Im} \Lambda) \rangle \propto (N \rho/k_0^3)^{-2/3}$ for $\rho/k_0^3 \gtrsim 1$ (see Fig.\ \ref{figminim}).

\subsection{Eigenvalue repulsion}
\label{repulsion}

\begin{figure*}
\includegraphics[width=0.88\textwidth]{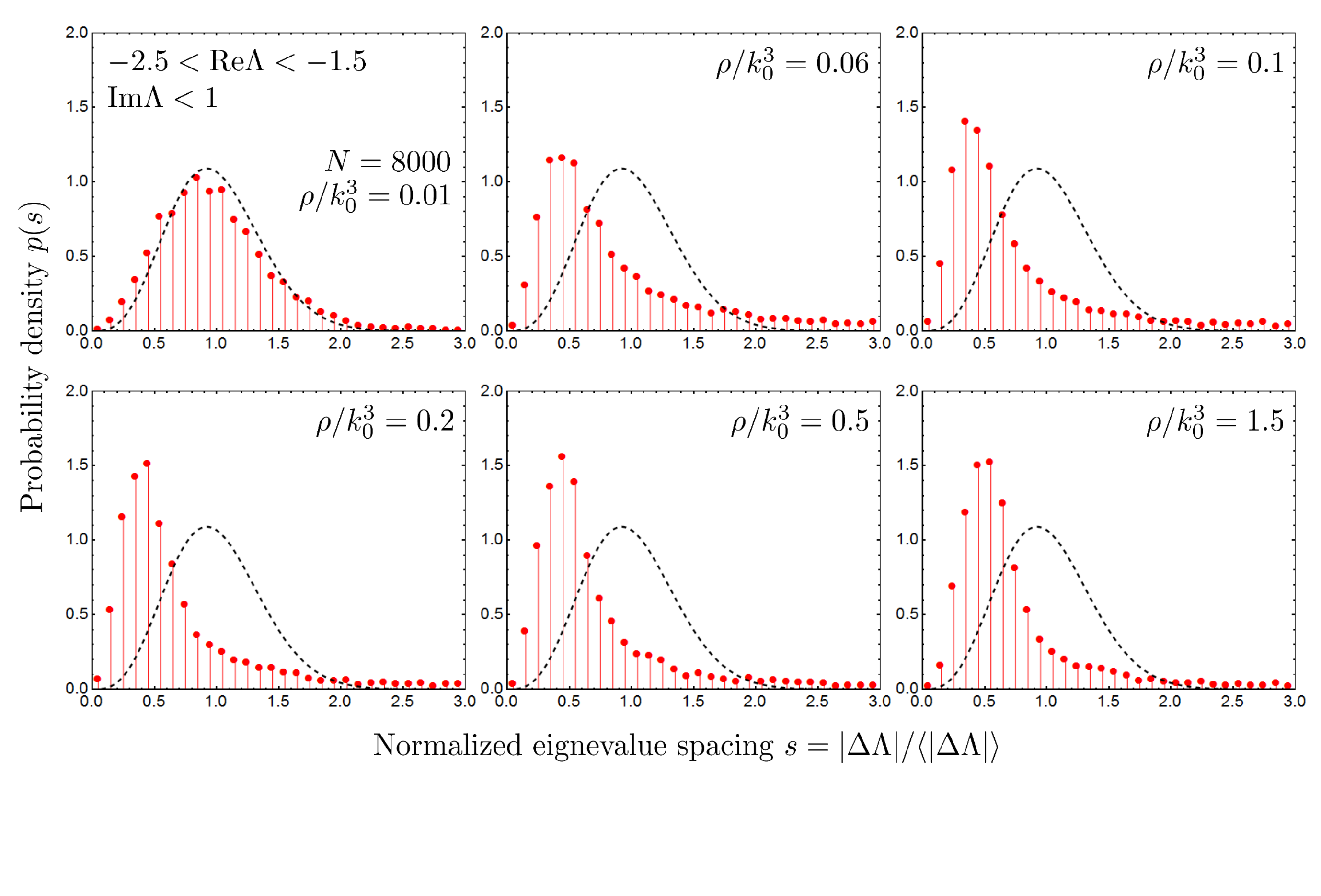}
\vspace*{-1.5cm}
\caption{Probability density of normalized spacings $s = |\Delta \Lambda|/\langle |\Delta \Lambda | \rangle$ of complex eigenvalues $\Lambda$ of the Green's matrix (2) at a strong magnetic field $\Delta = 10^3$. Only the eigenvalues with $-2.5 < \mathrm{Re} \Lambda -1.5$ and $\mathrm{Im} \Lambda < 1$ were taken into account. The probability density (\ref{ginibre}) expected for the Ginibre's random matrix ensemble is shown by the dashed line.}
\label{repulsion1}
\end{figure*}

\begin{figure*}
\includegraphics[width=0.88\textwidth]{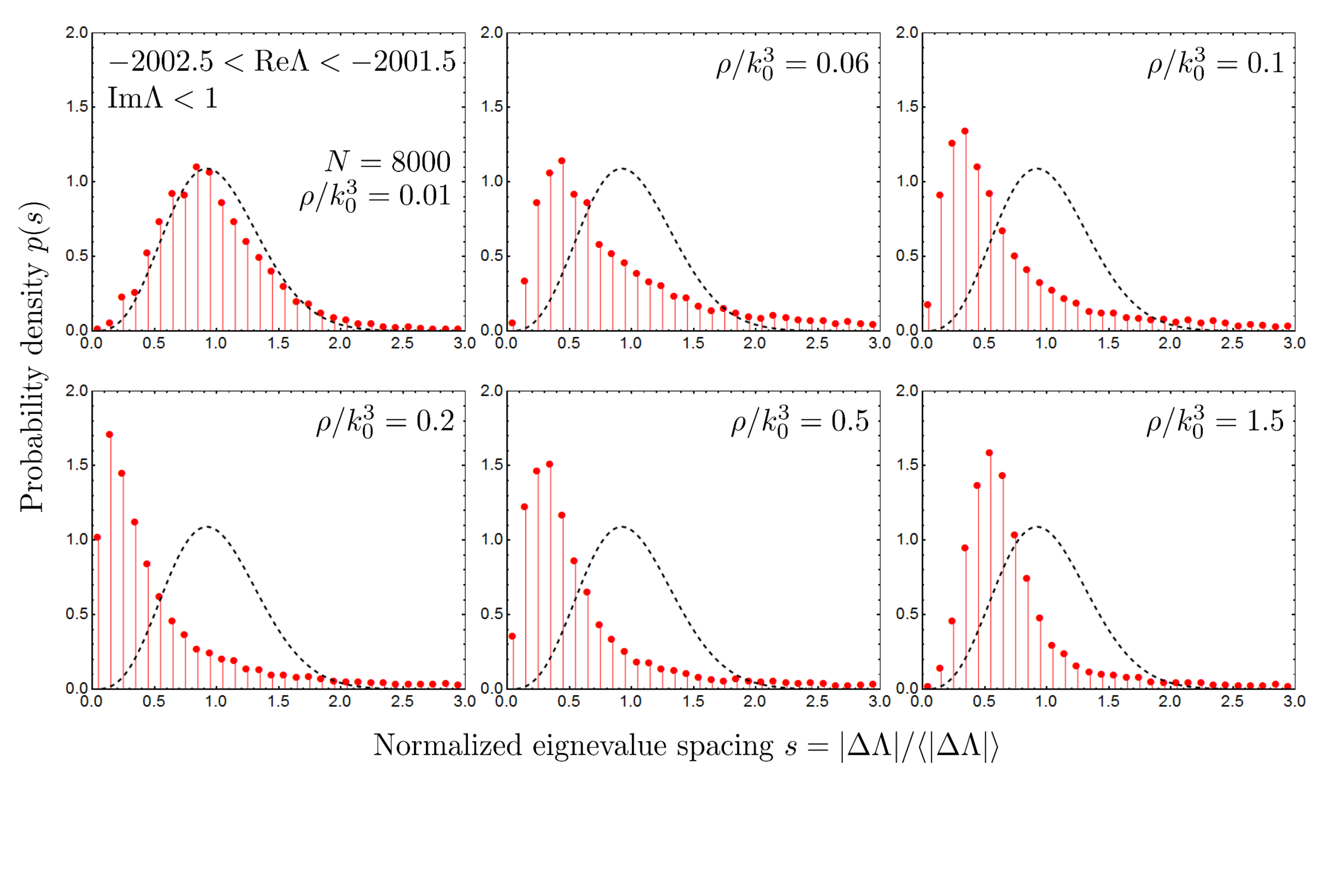}
\vspace*{-1.5cm}
\caption{Same as Fig.\ \ref{repulsion1} but for eigenvalues with $-2002.5 < \mathrm{Re} \Lambda < -2001.5$.
\vspace*{2cm}}

\label{repulsion2}
\end{figure*}

A well-known impact of localized states on the spectrum of a random matrix (or, more generally, of a disordered system) is the suppression of the so-called eigenvalue (or level) repulsion phenomenon. In brief, an Hermitian matrix (or a closed disordered system) with extended eigenstates is expected to exhibit eigenvalue (level) repulsion: the probability density function of spacings between nearest eigenvalues $\Lambda_i$ and $\Lambda_{i+1}$ \cite{foot3},
$|\Delta \Lambda| = |\Lambda_{i+1} - \Lambda_i|$, goes to zero for  $\Delta \Lambda \to 0$ \cite{mehta04x}. This is due to the mutual orthogonality of eigenvectors of an Hermitian matrix that forbids that two \textit{extended} eigenvectors correspond to the same eigenvalue. The appearance of localized states leads to the suppression of the eigenvalue repulsion because two states localized far from each other can now correspond to the same eigenvalue and $|\Delta \Lambda|$ can become arbitrary small.

\begin{figure}[t]
\includegraphics[width=0.9\columnwidth]{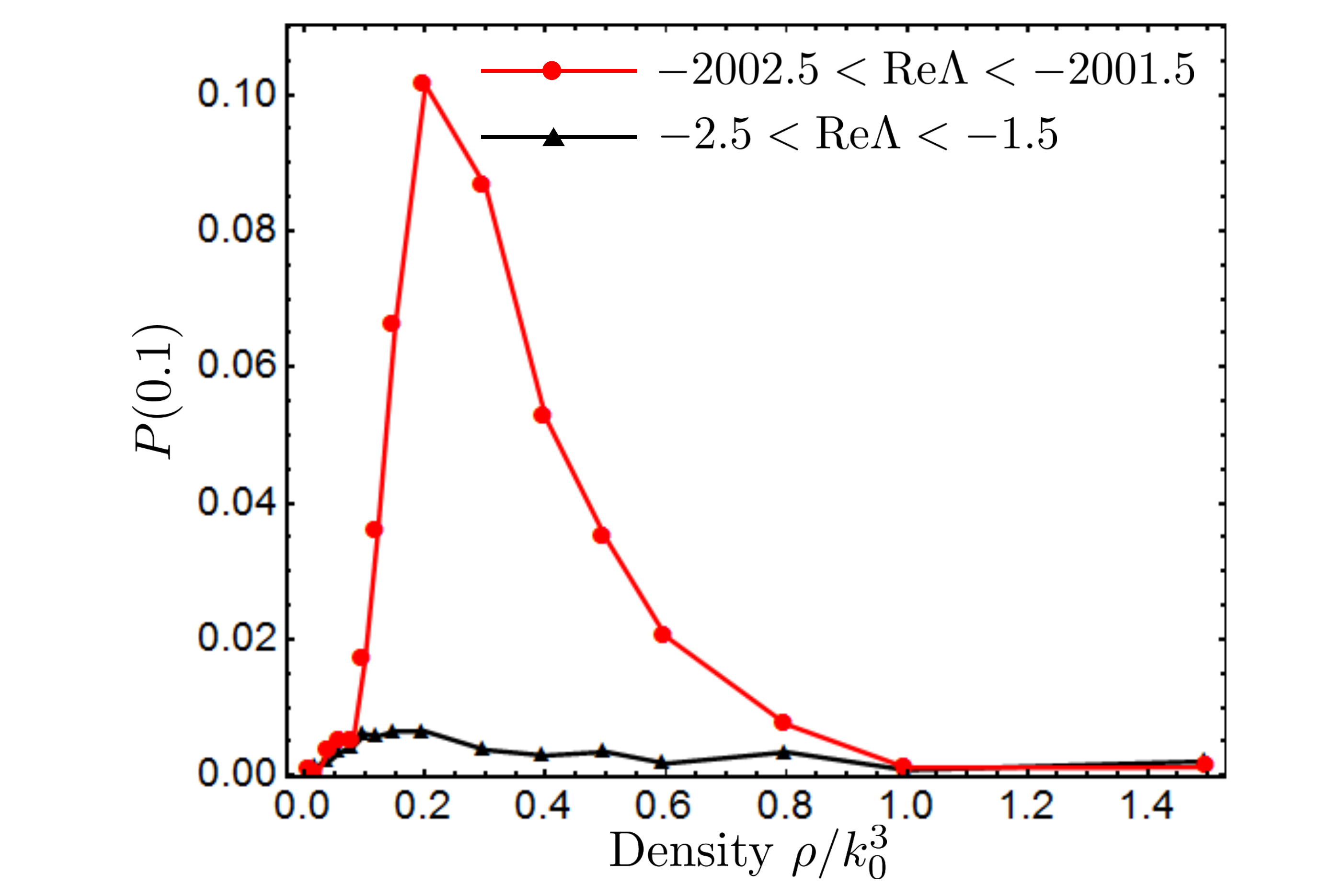}
\caption{The value of the cumulative distribution function of the nearest eigenvalue spacing $P(s)$ computed from the data of Figs.\ \ref{repulsion1} and \ref{repulsion2} is shown at a given small value of $s = 0.1$ as a function of atomic number density for two intervals of $\mathrm{Re} \Lambda$. }
\label{repulsion3}
\end{figure}

The concept of eigenvalue repulsion can be generalized to non-Hermitian matrices \cite{haake01x}. The eigenvalues $\Lambda$ are now complex but for each $\Lambda_i$ the nearest eigenvalue $\Lambda_j$ can still be identified as the eigenvalue that minimizes the distance between the two eigenvalues on the complex plane $|\Delta \Lambda| = |\Lambda_i - \Lambda_j|$. For the Ginibre's ensemble of random matrices \cite{foot4},
the probability density function of normalized eigenvalue spacings $s = |\Delta \Lambda|/\langle |\Delta \Lambda| \rangle$ is found to be \cite{haake01x}
\begin{eqnarray}
p(s) = \frac{3^4 \pi^2}{2^7} s^3 \exp\left( -\frac{3^2 \pi}{2^4} s^2 \right).
\label{ginibre}
\end{eqnarray}
We see that $p(s) \to 0$ for $s \to 0$ which means that two complex eigenvalues avoid being too close to each other on the complex plane. Importantly, the mutual repulsion of eigenvalues is stronger than it would be for the Poissonian random process on the plane that would lead to $p(s) \propto s \exp(-\pi s^2/4)$ \cite{haake01x}.

We now compute the probability density of normalized nearest eigenvalue spacings $p(s)$ for the complex eigenvalues $\Lambda$ of the Green's matrix (2) and compare the results obtained for unit bands of $\mathrm{Re} \Lambda$ centered at $\mathrm{Re} \Lambda = -2$ (where no localization transition is expected according to Fig.\ 2 and Fig.\ \ref{figiprcenter}) and at $\mathrm{Re} \Lambda = -2002$ (where we do expect a localization transition). To avoid the influence of irrelevant superradiant states with $\mathrm{Im} \Lambda > 1$, we limit our analysis to eigenvalues with $\mathrm{Im} \Lambda < 1$. As can be seen from Figs.\ \ref{repulsion1} and \ref{repulsion2}, at a low density $\rho/k_0^3 = 0.01$ the probability density $p(s)$ is close to the behavior predicted by Eq.\ (\ref{ginibre}) independent of the frequency interval under consideration. This can be understood from the fact that at small densities the atomic positions $\vec{r}_i$ are far apart so that most of the elements of the Green's matrix (2) become effectively uncorrelated due to the large phases $k_0 |\vec{r}_i - \vec{r}_j| \gg 2 \pi$. Statistical properties of the ensemble of Green's matrices thus approach those of the Ginibre's ensemble for which Eq.\ (\ref{ginibre}) was derived. In particular, the eigenvalue repulsion is clearly observed in both frequency intervals. At larger densities, however, the behaviors of $p(s)$ computed for $\mathrm{Re} \Lambda$ around $-2$ and around $-2002$ differ significantly. As we see from Fig.\ \ref{repulsion1}, for $\mathrm{Re} \Lambda$ around $-2$ where no localization transition is expected, $p(s)$ attains a roughly universal shape already for $\rho/k_0^3 \sim 0.1$. This shape hardly changes when the density is further increased. In contrast, for $\mathrm{Re} \Lambda$ around $-2002$ where a localization transition takes place, we clearly observe that $p(s)$ acquires an increasingly important weight at small $s$ when the density is increased up to $\rho/k_0^3 \approx 0.2$ (see Fig.\ \ref{repulsion2}). We interpret this increase of $p(s)$ for small $s$ as a suppression of eigenvalue repulsion due to the appearance of localized states. Further increase of density restores level repulsion signaling that the system leaves the localized regime and returns to the situation in which the eigenstates are extended. This picture becomes even more obvious if one looks at the cumulative distribution function $P(s)$, i.e. at the probability for the normalized nearest eigenvalue spacing to fall below a given $s$,
\begin{eqnarray}
P(s) = \int\limits_0^s p(s') \mathrm{d} s'.
\label{integrated}
\end{eqnarray}
In Fig.\ \ref{repulsion3} we show $P(s)$ for a small value of $s = 0.1$ as a function of atomic number density. The difference between the two intervals of $\mathrm{Re} \Lambda$ under study as well as an important suppression of eigenvalue repulsion leading to a more than ten-fold increase of $P(0.1)$ around $\rho/k_0^3 \approx 0.2$ are obvious.

\section{Remarks on light scattering by atoms in a magnetic field}
\label{remarks}

The results obtained in the main text of the Letter follow from the analysis of the exact $3N \times 3N$ Green's matrix (2) that take into account all the complex physical phenomena that take place in light scattering by atoms and relies on quite a limited number of reasonable approximations (see the main text and Refs.\ [17], [28], [29]). However, when trying to understand these results qualitatively, a number of legitimate questions may arise concerning the microscopic details of physical processes at work and their interplay. In this section, we provide elements of answers to a couple of such questions that we anticipate.

One might argue that given the decoupling of transitions corresponding to different $m = 0$, $\pm 1$ in a strong magnetic field,---the decoupling that is rigorously confirmed by the existence of the effective scalar Green's matrix (3) that describes scattering on each transition separately from the others,---light may escape from the medium along some special directions in which it has ``wrong'' polarization and cannot be scattered by the transition corresponding to its frequency. Indeed, imagine a photon emitted by an atomic excited state corresponding, say, to $m = 1$. The photon will be resonant with the same transition of the next atom on its way, but can be scattered by this atoms only if its helicity has a projection on the direction of the dipole moment of the atomic transition. The latter projection depending on the direction in which the photon propagates, a special direction will exist for which scattering vanishes, and the photon will be able to propagate ballistically until it reaches a boundary of the atomic cloud and escapes to the free space. Such ``escape channels'' induced by the magnetic field would decrease the lifetime of collective states in the atomic cloud and should not favor localized states. How can this picture be reconciled with our main conclusion about the appearance of localized states in the magnetic field?

\begin{figure}
\includegraphics[width=0.9\columnwidth]{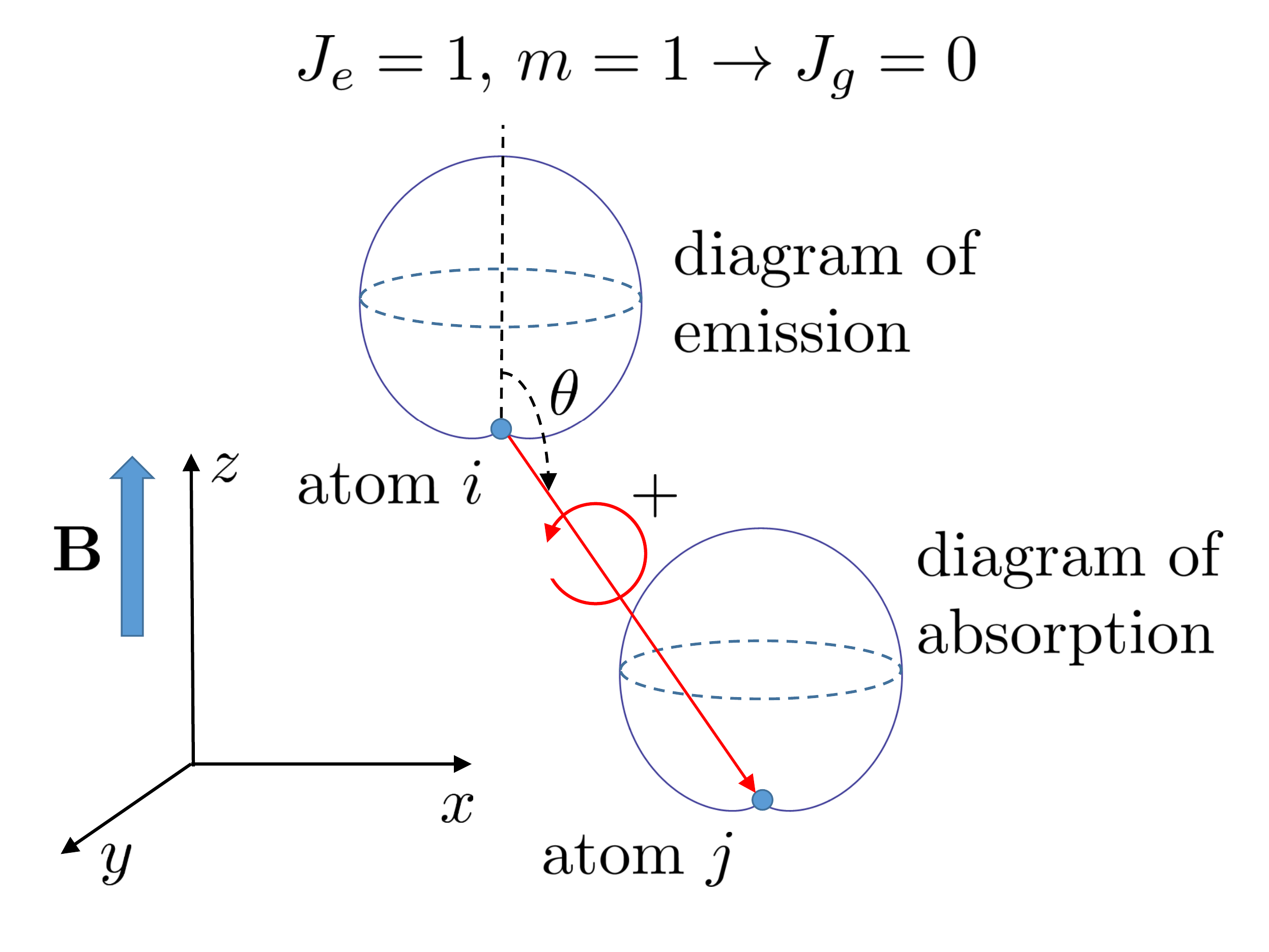}
\caption{Illustration of the processes of emission of a photon with right helicity on the transition $J_e = 1$, $m = 1 \to J_g = 0$ of the atom $i$ and its absorption on the transition $J_g = 0 \to J_e = 1$, $m = 1$ of the atom $j$. For angles $\theta$ close to $\pi$, the probability of absorption is small corresponding to the long scattering mean free path of the photon in a medium containing many atoms. But the probability of emission of a photon in such a direction is small as well so that such rare weakly scattering photons do not play a role in the statistical analysis that we present in the main text of the Letter.}
\label{photon}
\end{figure}

To answer this question, let us consider the problem a little bit more rigorously. The spontaneous decay of an initially excited atom $i$ (initial state $J_e = 1$, $m$) leads to the emission of a photon propagating along a direction determined by angles $\theta$, $\varphi$ of the spherical coordinate system (see Fig.\ \ref{photon}). To describe the polarization properties of radiation we will use the so-called spiral basis although the subsequent discussion can be repeated for the linear basis as well. The unit vectors of the spiral basis are denoted by $\vec{e}_-'$, $\vec{e}_0'$ and $\vec{e}_+'$. $\vec{e}_{\pm}'$ correspond to the right ($+$) or left ($-$) helicity of the photon whereas  $\vec{e}_0'$ is parallel to the direction of its propagation. In the rotating wave approximation the probability of emission of a photon with polarization $\vec{e}$ is determined by the scalar product of $\vec{e}^*$ and the dipole moment of transition $\vec{d}_{g_i e_{im}}$ \cite{cohen92x}. The three atomic transitions from excited states $J_e = 1$, $m = -1$; $J_e = 1$, $m = 0$; $J_e = 1$, $m = 1$ to the ground state $J_g = 0$ are characterized by three different vectors of the transition dipole moment $\vec{d}_{g_i e_{im}}$. The latter are parallel to unit vectors of the cyclic coordinate system $\vec{e}_{-}$, $\vec{e}_0$ and $\vec{e}_{+}$, respectively, with the quantization axis $z$ chosen parallel to the external magnetic field $\vec{B}$.

Because for a given atomic transition the probability of emission and of subsequent interaction of a photon with other atoms is determined by projections of spiral unit vectors on cyclic ones, a relation between the two bases is necessary to proceed. Such a relation can be found, for example, in Ref.\ \cite{varsh88x} and involves the direction of propagation of the photon determined by the angles $\theta$, $\varphi$:
\begin{eqnarray}
\vec{e}_+ &=& \frac12 \left[ \vec{e}_+' (1 + \cos \theta) -
\sqrt{2} \vec{e}_0' \sin \theta
\right. \nonumber \\
&+&
\left. \vphantom{\sqrt{2}} \vec{e}_-' (1 - \cos \theta) \right] \exp(\mathrm{i} \varphi),
\label{eq1}
\\
\vec{e}_0 &=& \frac{1}{\sqrt{2}} \left[ \vec{e}_+' \sin \theta +
\sqrt{2} \vec{e}_0' \cos \theta -
\vec{e}_-' \sin \theta \right],
\label{eq2}
\\
\vec{e}_- &=& \frac12 \left[ \vec{e}_+' (1 - \cos \theta) +
\sqrt{2} \vec{e}_0' \sin \theta
\right. \nonumber \\
&+&
\left. \vphantom{\sqrt{2}} \vec{e}_-' (1 + \cos \theta) \right] \exp(-\mathrm{i} \varphi).
\label{eq3}
\end{eqnarray}
Projecting cyclic unit vectors onto spiral ones we can determine the angular dependence of the probability amplitude of emission of a photon with a given helicity.

To start with, consider a photon emitted on the transition $J_e = 1$, $m = 1 \to J_g = 0$. In this case, the transition dipole moment $\vec{d}_{g_i e_{im}}$ is parallel to $\vec{e}_+$. From Eq.\ (\ref{eq1}) we see that probability amplitude $A_{+}$ for the photon to have right helicity is proportional to $\frac12 (1 + \cos \theta) \exp(\mathrm{i} \varphi)$ whereas the probability amplitude of having left helicity $A_{-}$ will be $\frac12 (1 - \cos \theta) \exp(\mathrm{i} \varphi)$. Photons with different helicities have different mean free paths. To estimate the latter, let us consider the probability $B_{\pm}$ of absorption of these photons by a second atom (atom $j$ in Fig.\ \ref{photon}) initially in the ground state $J_g = 0$. We assume that the photons are resonant with the transition $J_g = 0 \to J_e = 1$, $m = 1$ and neglect the existence of two other transitions which is justified in a strong magnetic field when the three transitions corresponding to different $m$ have very different frequencies.

Similarly to the emission probability amplitude, the absorption probability amplitude for a photon of right (left) helicity is determined by the scalar product of the unit vector $\vec{e}_{+}$ ($\vec{e}_{-}$) and the dipole moment $\vec{d}_{e_{jm} g_j}$. Because $\langle Jm | \hat{\vec{D}}_j | J'm' \rangle = \langle J'm' | \hat{\vec{D}}_j | Jm \rangle^*$, the required scalar product can be calculated using the complex conjugate of Eqs.\ (\ref{eq1}--\ref{eq3}). This yields the absorption probability amplitudes $B_+ = \frac12 (1+\cos \theta) \exp(-\mathrm{i} \varphi)$ and $B_{-} = \frac12 (1-\cos \theta) \exp(-\mathrm{i} \varphi)$ for photons of right and left helicities, respectively. On the one hand, these absorption amplitudes are smaller than those in the absence of magnetic field, when the photon emitted by the first atom is resonant with all three transitions of the second one. Moreover, for certain combinations of helicity and propagation direction (e.g., for the right helicity and $\theta = \pi$ or for the left helicity and $\theta = 0$), the absorption amplitude vanishes which corresponds to propagation without scattering and an infinitely large scattering mean free path for photons of these precise helicities in these precise directions. However, on the other hand, for the particular directions in which the probability amplitude of photon absorption by the atom $j$, $B_{\pm}$, vanishes, the probability amplitude of photon emission by the atom $i$, $A_{\pm}$, vanishes as well. In other words, there exist indeed certain combinations of helicity and propagation direction in which scattering is absent such that the corresponding photon would leave the atomic medium ballistically. However, the probability of emission of such a photon by an atomic excited state is zero and these particular combinations of helicity and propagation direction cannot serve as decay channels for the excited atomic states.

An extension of the above reasoning to photons emitted not exactly but close to the ``critical'' directions identified above shows that the scattering mean free path of these photons will be large but finite, whereas the probability of their emission will be small. As a result, they do not have any special influence on the quantities studied in the main text of the Letter (although, as we will see below, they are fully taken into account in our analysis). For example, no atomic states with anomalously short lifetimes appear due to these photons as can be verified by comparing eigenvalues of the random Green's matrix (2) with and without the external magnetic field $\vec{B}$. Moreover, the average lifetime of collective atomic states remains equal to $\Gamma_0/2$ (which follows from $\langle \mathrm{Im} \Lambda \rangle = 1$) independent of $\vec{B}$.

The arguments presented above can be repeated for the two other transitions with the conclusion remaining exactly the same. For a photon emitted on the transition $J_e = 1$, $m = -1 \to J_g = 0$, for example, the angular dependencies of emission and absorption probability amplitudes are given by the same equations as for the  photon emitted on the $J_e = 1$, $m = 1 \to J_g = 0$ transition with right and left helicities interchanged. For a photon emitted on the transition $J_e = 1$, $m = 0 \to J_g = 0$ the emission and absorption amplitudes are proportional to $\pm \sin \theta/\sqrt{2}$ for the right ($+$) and left ($-$) helicities, respectively. Once again, the directions $\theta = 0$, $\pi$ with a vanishing absorption probability correspond to a vanishing emission probability as well. We believe that the difference in the dependencies of emission/absorption probabilities for different helicities on the propagation direction for $m = 0$, $\pm 1$ may be at the origin of the peculiar difference observed between the localized states appearing near $\mathrm{Re} \Lambda = \pm 2 \Delta$ (states resonant with the transitions $J_e = 1$, $m = \mp 1 \to J_g = 0$) and extended states near $\mathrm{Re} \Lambda =0$ (states resonant with the transitions $J_e = 1$, $m = 0 \to J_g = 0$) although more work is required to establish a precise link between the two phenomena.

Finally, we would like to note that the analysis presented above allows for calculating the total probability amplitude $C_m$ for an atom to get excited by the photon emitted by another atom (see Fig.\ \ref{photon}). Important for us is the dependence of this amplitude on the angles $\theta$ and $\varphi$ that can be obtained by multiplying the probability of emission $A$ by the probability of absorption $B$ and then summing over the two helicities. We obtain
\begin{eqnarray}
C_{\pm} &\propto& \frac14 (1+\cos\theta)^2 + \frac14 (1-\cos\theta)^2
\nonumber \\
&=& 1 - \frac12 \sin^2 \theta,
\label{exp1}
\\
C_0 &\propto& \sin^2 \theta.
\label{exp2}
\end{eqnarray}
These results coincide with those that can be obtained from the Green's matrix (2). Indeed, for two atoms $i$ and $j$ separated by a distance $r_{ij}$  much exceeding the wavelength, the probability amplitude of photon scattering on the resonant transition following from Eq.\ (2) is
\begin{eqnarray}
C_{m} \propto |d_{g_j e_{jm}}^m|^2 \left(
1 - \frac{|r_{ij}^{m}|^2}{r_{ij}^2} \right).
\label{exp3}
\end{eqnarray}
Here $d_{g_j e_{jm}}^m$ and $r_{ij}^{m}$ are the cyclic components of the vectors $\vec{d}_{g_j e_{jm}}$ and $\vec{r}_{ij}$, respectively. Equation (\ref{exp3}) reduces to $C_{\pm} \propto 1 - \frac12 \sin^2 \theta$ and $C_{0} \propto \sin^2 \theta$ and coincides exactly with the predictions of Eqs.\ (\ref{exp1}) and (\ref{exp2}).

\section*{Conclusions}
\label{concl}

The analysis presented in section \ref{evidence} clearly shows that the localization transition discovered in the main text exhibits a number of features expected for the disorder-driven, Anderson localization transition. In particular, the minimum decay rate of quasi-modes decreases exponentially with sample size beyond a certain critical number density of atoms (section \ref{decay}) and the mutual repulsion of complex eigenvalues is suppressed (section \ref{repulsion}). In addition, the localized states under study can be clearly distinguished from subradiant states localized on pairs of closely located atoms due to very different values of the corresponding eigenvalues $\Lambda$ ($\mathrm{Re} \Lambda$ near the resonances $\mathrm{Re} \Lambda \simeq -2 m \Delta$ for the new discovered localized states but $\mathrm{Re} \Lambda$ far from resonances for the two-atom subradiant states, see section \ref{iprmaps}). Despite this, however, it would be premature to claim that the discovered transition is a standard Anderson transition because the role of cooperative phenomena and of dipole-dipole interactions between atoms still remains to be clarified.

In section \ref{remarks} we have demonstrated that escape channels---special combinations of polarization and propagation direction for which a photon is not scattered by the atoms,---do not affect the lifetime of the excited atomic states and are thus compatible with the localization transition discovered in the main text of the Letter. This is due to the relation that exists between the emission and absorption diagrams of two-level atoms in the magnetic field. The latter relation leads to a vanishing emission probability for photons that would have a vanishing absorption probability. Therefore, such photons are not emitted by the excited atomic states and thus the escape channels cannot serve as efficient decay channels for the excited atomic states.


\begin{thebibliography}{999}
\bibitem{foot1}
We will use the terms ``eigenvector'', ``eigenstate'', ``state'' and ``quasi-mode'' interchangeably depending on the context.
\bibitem{foot3}
The real eigenvalues $\Lambda_i$ are assumed to be ordered.
\bibitem{foot4}
Ginibre's ensemble is composed of large ($N \to \infty$) random matrices with independent identically distributed complex elements having zero mean and Gaussian statistics.
\end{thebibliography}
\end{document}